\documentclass[aps,prl,twocolumn,showpacs,
               preprintnumbers,amsmath,amssymb,floatfix]{revtex4}

\usepackage{graphicx}
\usepackage{bm}
\usepackage{amssymb}

\newcommand{\bd}{../../../bib}

\newcommand{\avg}[1]{\ensuremath{\langle\ensuremath{#1}\rangle}}

\newcommand{\fig}{\begin{figure}}
\newcommand{\efig}{\end{figure}}

\newcommand{\figref}[1]{Fig. \ref{#1}}

\newcommand{\subfigsref}[3]{Fig. \ref{#1}(#2--#3)}

\newcommand{\subfigref}[2]{Fig. \ref{#1}(#2)}
\newcommand{\Subfigref}[2]{Figure \ref{#1}(#2)}
\newcommand{\Table}{\begin{table}}  
\newcommand{\etable}{\end{table}}  
\newcommand{\eq}{\begin{equation}}  
\newcommand{\eeq}{\end{equation}}  
\newcommand{\eqa}{\begin{eqnarray}}  
\newcommand{\eeqa}{\end{eqnarray}}  

\begin{document}

\preprint{rev1.03}

\title{First-order phase transition and the equation of state 
in a 2D granular fluid}
\author{M. D. Shattuck}
\email[]{shattuck@ccny.cuny.edu}
\homepage{http://gibbs.engr.ccny.cuny.edu}
\affiliation{Benjamin Levich Institute and Physics Department,
The City College of the City University of New York
140th and Convent Ave., New York, NY  10031}

\date{\today}

\begin{abstract}
We present experimental evidence for a first-order freezing/melting
phase transition in a nonequilibrium system --- an oscillated
two-dimensional isobaric granular fluid.  The steady-state transition
occurs between a gas and a crystal and is characterized by a
discontinuous change in both density and temperature. It is suppressed
if the number of particles is incommensurate with the cell size, shows
rate-dependent hysteresis, and obeys the Lindemann criterion for
melting.  Further, the measured equation of state both above and below
the phase transition compares well with theory.
\end{abstract}


\maketitle

{\em Introduction:} Granular materials are fascinating systems, which
have tremendous technological importance and numerous applications to
natural systems.  They also represent a serious challenge to
statistical physics as an extreme example of a system far from
equilibrium.  As for any macroscopic system, the total number of modes
in a collection of grains is on the order of Avogadro's number $N_A$.
However, in granular systems a very small number of modes,
specifically the translational and rotational mode of the $N$
macroscopic grains which make up the system, can be preferentially
excited by external forces like moving walls.  The energy in these
$\sim N$ modes can be many orders of magnitude above the average
energy per mode (i.e., the temperature) for the remaining $N_A$ modes.
The study of the relaxation of this extremely far-from-equilibrium
system toward equilibrium is the challenge of granular statistical
mechanics.  In this paper, we experimentally examine the
nonequilibrium steady-state (NESS) created by a balance between
relaxation (dissipation) and injection of energy (heating), and
specifically, a first-order melting/freezing phase transition and the
equation of state in a two-dimensional (2D) inelastic hard sphere
system heated (excited) from below under isobaric conditions.

The elastic hard sphere system is the simplest system which undergoes
a first-order phase transition\cite{alder62,hoover68}.  This
transition has been seen in simulation \cite{alder68b} and experiments
in colloids \cite{rutgers96}.  However, in these systems energy is
conserved, and the concept of free energy is well defined.  In
dissipative systems, such as inelastic hard spheres or granular
systems, the concept of free energy is not established
\cite{kadanoff99}, even in a NESS, since there is a constant flow
of energy through the system.  The idea of applying thermodynamics
to a NESS is at the forefront of current statistical physics
\cite{Ruelle99}.  The demonstration of experimental systems which
undergo first-order phase transitions under NESS conditions points to
the universality of entropy-production and free-energy concepts even
in the absence of energy conservation.

\fig
\includegraphics[width=3.375in]{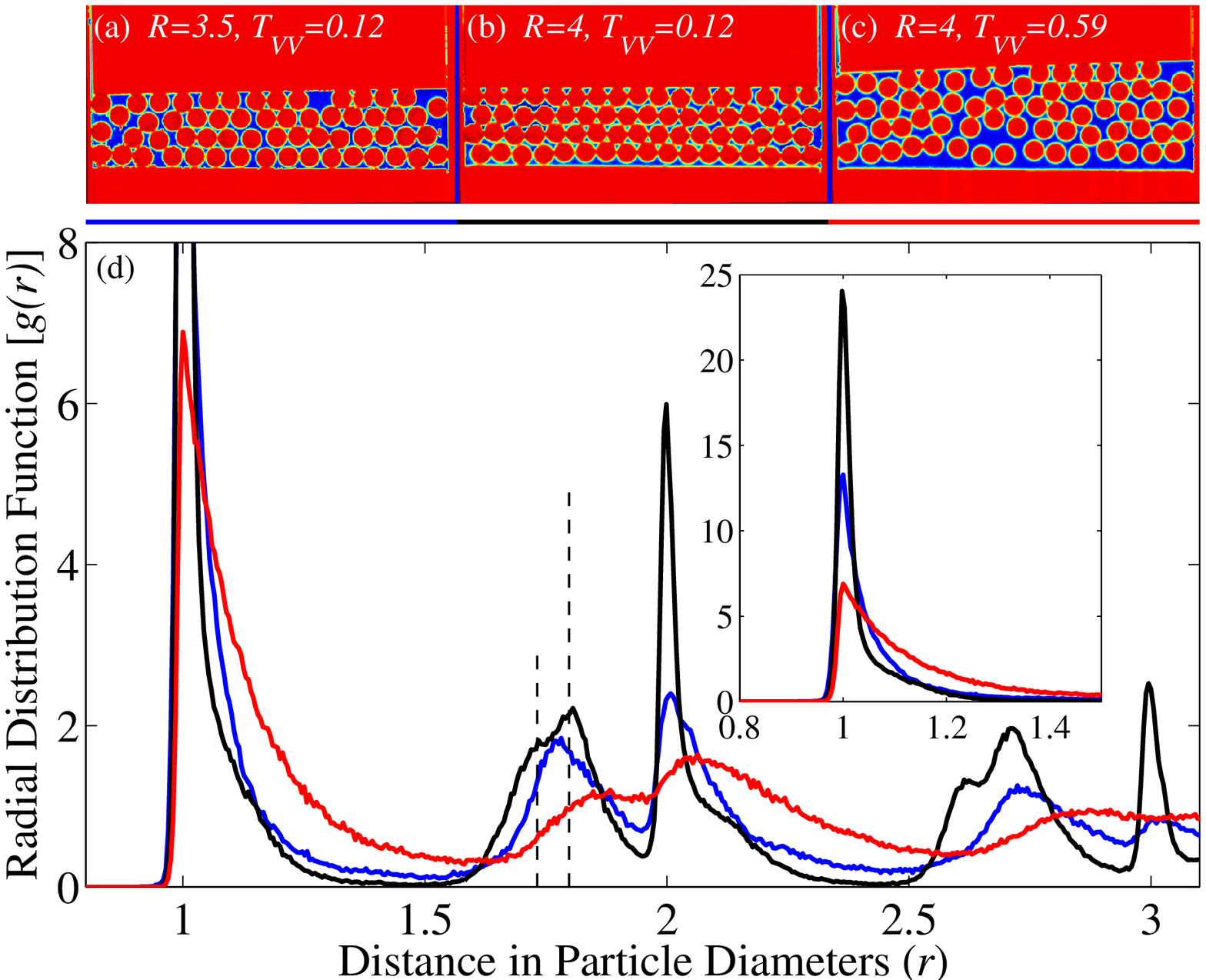}
\caption{(a)-(c) Photographs of a 2D granular layer, with freely
  floating (isobaric) weight: (a) $R=3.5$, $\Gamma=3.88$,
  $T_{VV}=0.12$, disordered with crystalline regions.  (b) $R=4$,
  $\Gamma=7.90$, $T_{VV}=0.12$, crystal.  (c) $R=4$, $\Gamma=8.10$,
  $T_{VV}=0.59$, gas.  (d) The radial distribution function for states
  (a)-(c). The line style is indicated below each image.  The vertical
  lines show next-nearest-neighbor distances.  The short line is
  $\sqrt{3}D$ and the long line is $\sqrt{3}$ times the average
  nearest-neighbor distance in (b).  The inset shows the first peak at
  full scale.}
\label{cellgr}
\efig

{\em Experiment:} We place $N$ ($26$--$85$) spherical stainless steel
ball bearings of diameter $D=3.175$ mm in a container 17.5 $D$ wide by
20 $D$ tall by 1 $D$ deep as shown in \subfigsref{cellgr}{a}{c}.  We
define the number of rows $R=N/17$, where 17 is the number of
particles to fill an entire row.  A thin plunger slides through a slot
in the bottom of the cell and oscillates sinusoidally to excite (heat)
the particles from below.  The driving is characterized by the
nondimensional maximum acceleration $\Gamma = A(2\pi f)^2/g$, where
$A$ is the maximum amplitude of the plunger, $f$ is the frequency and
$g$ is the acceleration of gravity.  A freely floating weight confines
the particles from the top, allowing the volume to fluctuate but
providing constant pressure conditions.  When $\Gamma$ is large the
granular system acts like a gas, as shown in \subfigref{cellgr}{c}.
When $\Gamma$ is small and $R$ is an integer a crystalline state
develops [see \subfigref{cellgr}{b}].  Using high-speed digital
photography we measure the positions of the plunger, the weight, and
all of the particles in the cell with a relative accuracy of 0.2\% of
$D$ or approximately $6 \mu$m at a rate of 840 Hz. We track the
particles from frame to frame and assign a velocity to each one,
typically $\sim D/5$ per frame.

From these data we measure the volume, density, and granular
temperature or average kinetic energy per particle.  In a normal gas,
the temperature is isotropic and the kinetic energy in vertical
velocity is the same as that in horizontal velocity.  In a granular
system, due to dissipation, at least two temperatures are needed in
two dimensions, a horizontal temperature $T_{HH} = 1/2 m
\avg{(v_x-\avg{v_x})^2}$ and a vertical temperature $T_{VV} = 1/2 m
\avg{(v_y-\avg{v_y})^2}$, where $m$ is the mass of the particles and
$v_x$ and $v_y$ are the horizontal and vertical velocities
respectively.  We want to assign a single value to each temperature,
so we take the averages over all particles and time ($\sim 50$
periods).  The most obvious implementation of the averages $\avg{...}$
is in the inertial lab frame; however, this produces a finite
temperature even if the plunger, particles, and weight all move with
the same velocity.  Taking the averages in the center of mass frame of
the particles solves this problem, but the velocities are no longer
simply related to the collision velocities with the weight used in the
derivation of the equation of state.  We have chosen to evaluate the
averages in the frame of the weight.  This choice solves both
problems.  The ultimate solution is to allow the temperature, density,
and pressure to vary in space and time, and look at the space and time
dependent equation of state.  However, our current approach is much
simpler, and the experimental equation of state compares well with
standard kinetic theory \cite{jenkins85}.

We measure the average density, horizontal and vertical temperatures,
and radial distribution functions $g(r)$ for three different constant
pressures, for $R$ from 1.5 -- 5, and for $\Gamma$ from 0--30.  To
prepare the system initially in the densest state, we increase
$\Gamma$ to 30 and then slowly lower the acceleration to zero.
Then we alternate taking 1024 pictures with a small increase in
$\Gamma$ and a variable time delay (typically one second) until
$\Gamma=30$.  Then we repeat the same process while decreasing $\Gamma$
to zero.  Typical data runs are shown in \figref{Tnuvamp}.

\fig
\includegraphics[width=3.375in]{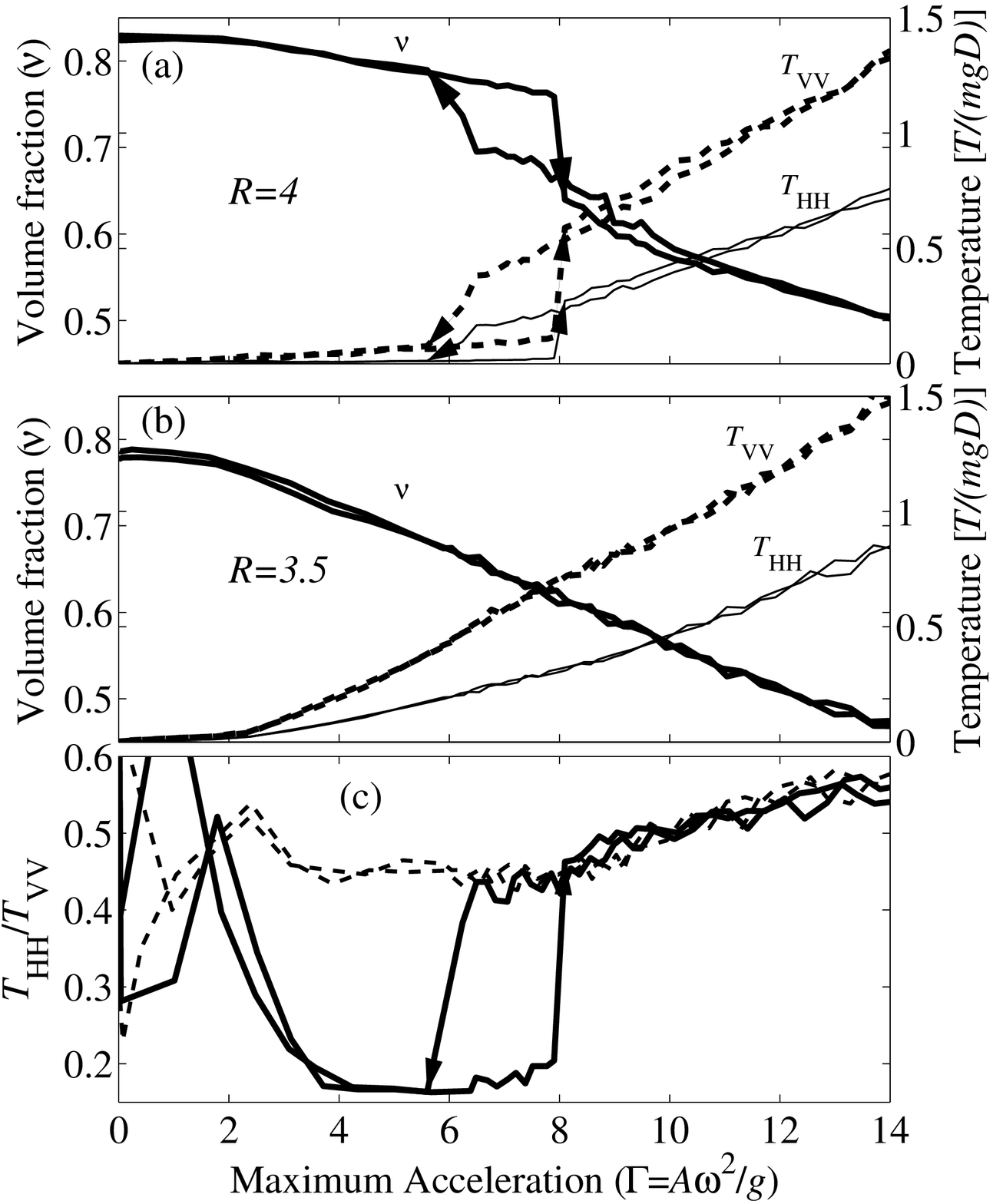}
\caption{(a)-(b) Plots of the volume fraction (thick) and average
  vertical (dashed) and horizontal (thin) temperatures as a function
  of increasing and decreasing $\Gamma$ at 50 Hz, under isobaric
  conditions.  (a) First-order phase transition for $R=4$, showing a
  discontinuities in density and temperature as well as
  hysteresis. (b) For $R=3.5$, the curves are smooth and
  non-hysteretic.  (c) Plot of the horizontal to vertical temperature
  ratio for $R=3.5$ (dashed) and $R=4$ (solid).}
\label{Tnuvamp}
\efig

{\em Results:} The behavior of the system is quite different depending
on the number of particles in the cell.  If $R$ is an integer then we
observe a first-order hysteretic phase transition as we change
$\Gamma$.  \Subfigref{Tnuvamp}{a} shows a typical example for $R = 4$.
Below $\Gamma = 1$, nothing happens --- the temperatures are zero and
the 2D volume fraction $\nu$ is constant.  As we increase $\Gamma$ for
$1 < \Gamma < 8$, $T_{VV}$ rises linearly.  Initially, $T_{HH}\simeq
T_{VV}/2$, but around $\Gamma=3$ the ratio drops and stays around
20\%, as shown in \subfigref{Tnuvamp}{c}. $\nu$ remains nearly
constant for $\Gamma<2$ and then falls linearly until $\Gamma = 8$.
At $\Gamma = 8$, there is an abrupt change in all of the measured
quantities.  $T_{VV}$ and $T_{HH}$ rise by factors of 5 and 10
respectively, and the ratio reaches nearly 50\%.  $\nu$ drops by more
than 15\%.  For $\Gamma > 8$, $T_{VV}$ and $T_{HH}$ continue to rise,
the ratio reaches 70\% by $\Gamma=30$ (not shown), and $\nu$ continues
to drop.  As $\Gamma$ is lowered, the process is reversed, but the
transition point shows 25\% hysteresis, occurring at $\Gamma = 6.1$.
The hysteresis is rate-dependent, and if the rate of change in
$\Gamma$ is increased by a factor of 100 the hysteresis is lost.
Similar behavior is seen for $R =$ 2, 3, and 5, although the size of
the jumps decrease as $R$ is decreased.  For $R = 3.5$ the situation
is quite different, as shown in \subfigref{Tnuvamp}{b}.  All of the
measured quantities change continuously as $\Gamma$ is changed, and no
hysteresis is observed.  The ratio of $T_{VV}$ to $T_{HH}$ stays near
50\% until $\Gamma = 1$.  Similar behavior is seen for $R =$ 1.5, 2.5,
and 4.5.

$g(r)$ is shown in \subfigref{cellgr}{d} for the $R=4$ states just
before ($\Gamma=7.90$) and just after ($\Gamma=8.10$) the transition.
While $\Gamma$ only changes by 2.5\% $g(r)$ changes dramatically.
$g(r)$ for the gas state is typical for a hard sphere fluid near a
freezing transition, including the broad peak just below $r=2D$ that is
a precursor for freezing \cite{truskett98}.  $g(r)$ for the
crystalline state is typical of an expanded hard sphere crystal at
nonzero temperature, showing sharp peaks at $r=D, 2D,$ and $3D$.
There is a big peak at the next-nearest neighbor distance
$r=\sqrt{3}D$ with a splitting due to the fact that the average
nearest neighbor distance is greater than $D$ at a nonzero
temperature.  $g(r)$ for $R=3.5$ at $\Gamma=8$ is not shown but is
clearly fluid-like.  Even at $\Gamma=3.88$ where the $T_{VV}$ is equal
to the crystal melting temperature, $g(r)$ is intermediate between a
crystal and fluid, with small peaks at integer separations, as shown
in \subfigref{cellgr}{d}.  In the image [\subfigref{cellgr}{a}]
crystalline regions can be seen, but there are gaps, holes, and
dislocations.  Information from $g(r)$ is summarized in
\subfigsref{lind}{b}{c}.  In all of the crystallizing systems
particles are excluded from a band at $r=1.5D \pm 0.1D$ and the
average number of next nearest neighbors is near the maximum as shown
in \subfigsref{lind}{b}{c} . The integer system is clearly in a
crystalline state, but it is not clear if the half-integer system ever
reaches a crystalline state.  Another classic measure of melting is
the Lindemann criterion \cite{lindemann10}, in which the average
root-mean-squared particle displacements
$\gamma_m=\sqrt{\avg{(\vec{r}-\avg{\vec{r}})^2}}/D > \gamma_c$.
$\gamma_m$ is plotted against $\Gamma$ in \subfigref{lind}{a}.  We
find a value of $\gamma_c$ between 0.10--0.12 is consistent with all of
the integer data.  This also predicts that the half-integer systems
freeze, but at a lower $\Gamma$, around 4 for $R=3.5$.  However, the
predicted freezing temperatures for all data fall in a band between
the crystal freezing and melting temperatures ($T_{VV}=0.1$ -- $0.6$).
While the Lindemann criterion suggests that all of the systems freeze,
it says nothing definitive about the structure of the solid and the
radial distribution functions suggest that freezing is a continuous
process for the half-integer systems.  Recent work on elastic hard
spheres using density-functional theory by Both and Hong
\cite{both2001} predicts a critical melting temperature determined by
a constant $\mu_0$.  From our melting and freezing temperature we find
a range of $\mu_0$ between $10$--$60$ which is consistent with their
findings.

\fig
\includegraphics[width=3.375in]{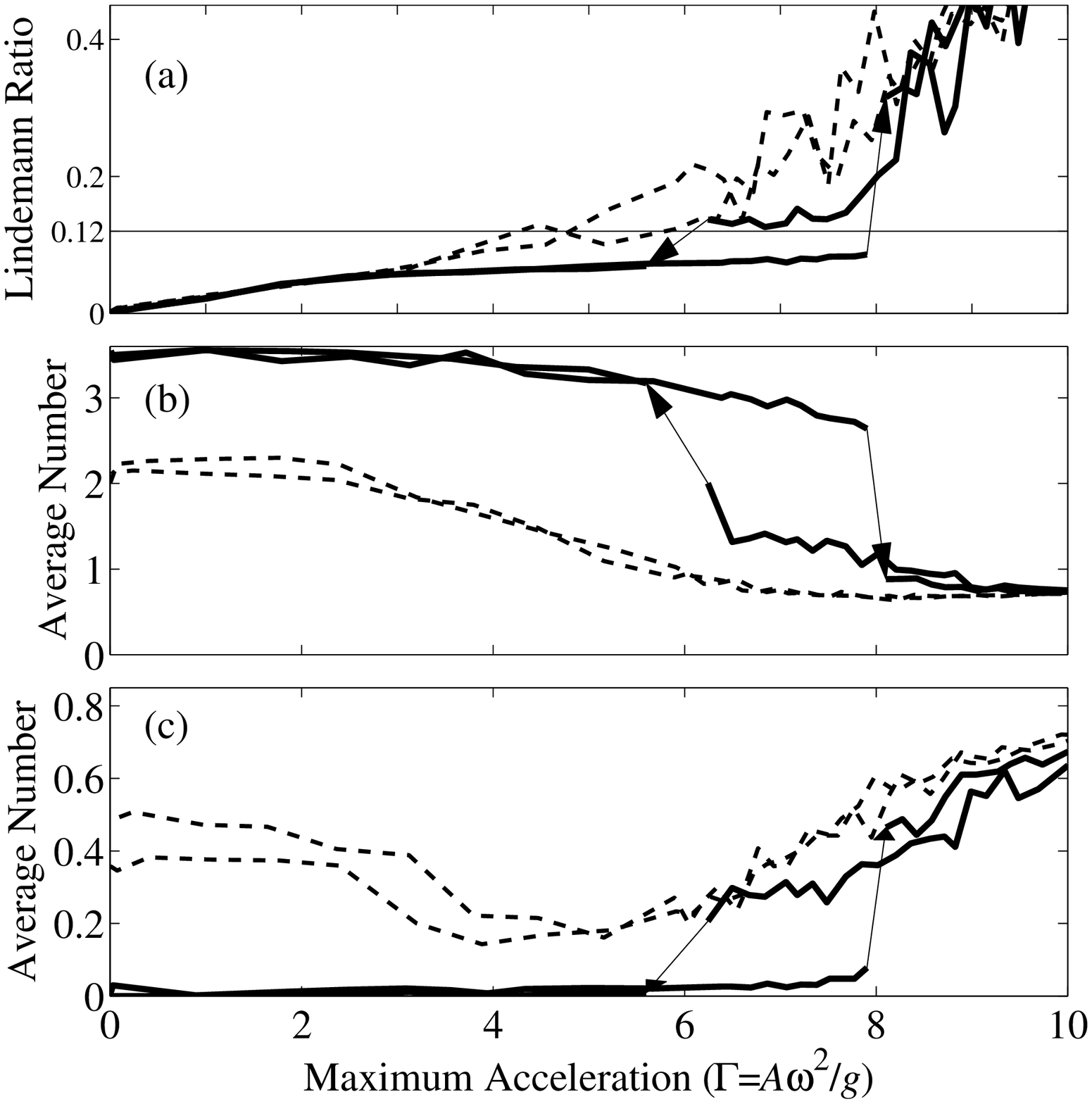}
\caption{Plots of the (a) Lindemann ratio and the average number of
  particles separated by (b) $\sqrt{3}D \pm 0.1D$ and (c) $1.5 D \pm
  0.1D$ versus $\Gamma$ for $R=4$ (thick) and $R=3.5$ (dashed).}
\label{lind}
\efig

To further exploit the analogy to thermodynamics, we have examined the
average equation of state for this system.  From $T_{VV}$, the average
number density $n=4\nu/(\pi D^2)$, and average pressure determined
from the mass of the weight and half of the mass of the particles, we
experimentally measure the compressibility factor $\chi$ from the
equation of state for this 2D granular fluid\cite{jenkins88}, $P = n
T_{VV} (1+\chi(\nu))$; $\chi=P/(n T_{VV}) -1$.  $\chi$ can be
theoretically related to $g(D)$, through the viral for the pressure
\cite{chapman70}.  In an inelastic hard sphere fluid
$\chi(\nu)=\alpha\nu g(D;\nu)\equiv\alpha G(\nu)$, where
$\alpha=(1+e)$ and $e$ is the coefficient of restitution.  For an
elastic hard sphere fluid $e=1$ and $\alpha=2$.  A number of forms for
$G$ are available in the literature (see \cite{luding2001} for a
recent discussion).  All are similar and we use the simple $G_T$
developed by Torquato \cite{torquato95}, which is an analytical fit to
molecular dynamics simulation at high $\nu$ and the Carnahan and
Starling \cite{carnahan69} geometric series approximation to the first
few viral coefficients at low $\nu$.

A plot of $\chi$ versus volume fraction is shown in \figref{chi} for
$R = 3.5$ and 4.  $\chi$ remains $\sim$~10\% of $2G_T$ up to $\nu=
0.55$, $R = 3.5$ and 4.  Above 0.55 the half-integer system drops
$\sim$~10\% below $2G_T$, but both $\chi$ and its derivative are
continuous functions of $\nu$ up to $\chi \simeq 200$.  Above this
value, $\Gamma$ is near 1, the temperatures are almost zero, and
$\chi$ strongly diverges.  This behavior is expected, since our
equation of state completely neglects the elastic forces between the
particles.  When the temperature goes to zero the weight is held up by
elastic energy, not kinetic energy.  However, the point at which this
occurs is a very high density in which the pressure due to the density
of the gas through the compressibility factor is 200 times the value
of a dilute gas.  The behavior above $\nu = 0.6$ is quite different
for the integer systems.  As $\nu$ increases there is a jump in both
$\chi$ and $\nu$ signaling the phase transition.  Above this volume
fraction the system is in an expanded crystalline state like that
shown in \subfigref{cellgr}{b}.  The size and position of the gap
depend on a number of parameters, including the history, number of
rows, and confining pressure.  In the crystalline state $\chi$
increases until $\Gamma$ is nearly one and the maximum recorded
packing fraction is reached.  At this point $\chi$ diverges strongly
as elastic forces become important.  As with the half-integer case, the
granular temperature must be very low and the density must be very
close to the maximum before elastic consideration are important.  This
suggests that continuum theories based solely on kinetic
considerations may have a range of applicability up to very high
density and very low temperatures.

To see the comparison between $\chi$ and $G_T$ more clearly, the inset
of \figref{chi} plots $\chi/G_T$ which should be 2 for elastic hard
spheres and less for inelastic hard spheres.  We can also measure $G$
from the experimental $g(D;\nu)$ and compare that to the experimental
value of $\chi$.  $G_{exp}$ determined in this manner is only accurate
for $\nu<0.8$; above this value the peak near $D$ begins to split due
to small imperfections in the crystal lattice, and the peak height
stops growing.  From $G_{exp}$ we obtain a fully experimental measure
of $\alpha$.  The result is also shown on the inset of \figref{chi}.
$\alpha_{exp}$ is consistent with our direct measures of $e$ from
particle tracking.  We find values of $e$ over a broad ranging from
0.3 to 1.3, and a general trend of lower values for higher collision
velocities.  Values of $e>1$ are possible since we cannot measure
particle spin.

\fig
\includegraphics[width=3.375in]{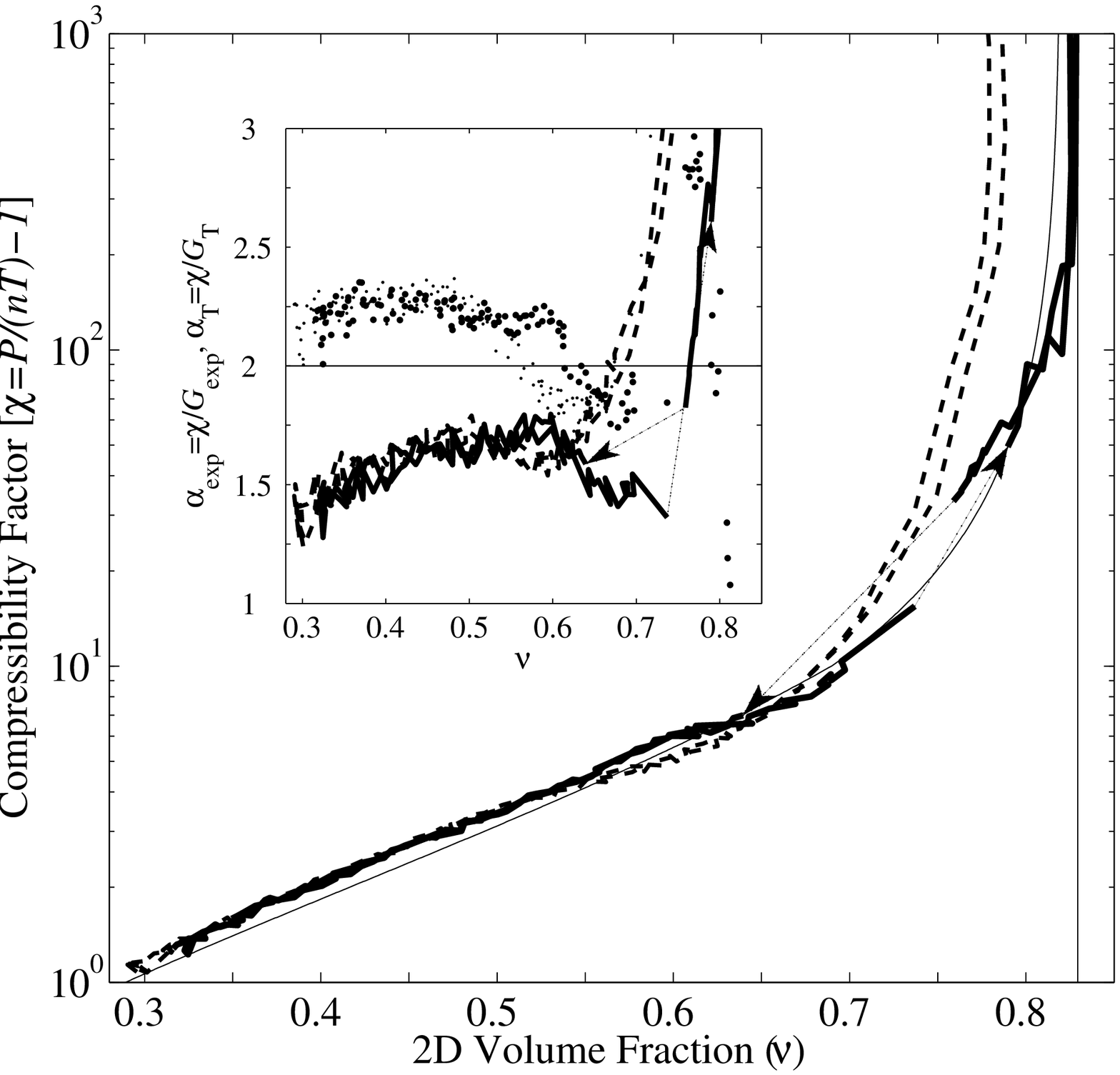}
\caption{Log-linear plot of the compressibility factor as a function
  of volume fraction, for $R=3.5$ (dashed), $R=4$ (thick), and $2G_{T}$
  (thin, see text). The dashed ($R=3.5$) and thick ($R=4$) curves in
  the inset show an experimental measurement of $\alpha_{exp}$ (see
  text) as a function of volume fraction.  $\alpha=2$ for elastic hard
  spheres.  The large dots ($R=4$) and small dots ($R=3.5$) show
  $\alpha_T$ using $G_T$.}
\label{chi}
\efig

{\em Discussion:} The first-order phase transition seen in the integer
number of rows is qualitatively different from that of an elastic hard
sphere system.  In such a system, just as in an ordinary gas, at the
phase transition there is a discontinuous change in the density,
but the temperature would be unchanged.  This is a unique feature in
granular systems since it suggests that there is the possibility of
steady states in which two phases co-exist, but at different nonzero
temperatures.  States like these have been seen experimentally
\cite{olafsen98} and in simulation \cite{meerson2003}.  This
coexistence does not violate thermodynamics since a steady state is
not the same as equilibrium, but it can have profound effects on
continuum theories since it predicts discontinuities in the
temperature field could exist in steady state without the need for
energy or particle flow.  That is, the system is mechanically stable
since there are two states with the same pressure but different
densities and internal energies.  This would also interfere with the
standard continuum assumption of near Maxwell-Boltzmann distributions,
since near the gas/crystal border two different velocity distributions
can co-exist.  Any spatial average would include some
of each distribution.  So while the equation of state data suggests
that theories based solely on kinetic consideration are applicable on
either side of the phase transition, some modification will be needed
to account for the discontinuities created by phase transitions.

\begin{acknowledgments}
This work was supported by The National Science Foundation, Math,
Physical Sciences Department of Materials Research under the Faculty
Early Career Development (CAREER) Program: DMR-0134837.
\end{acknowledgments}

\bibliography{\bd/newsand,\bd/stress,\bd/sound,\bd/silo,\bd/nagel,\bd/behringer,\bd/force,\bd/mri,local}

\end{document}